\documentclass[conference]{IEEEtran}

\usepackage{layouts}
\usepackage[utf8]{inputenc}
\usepackage{amsmath}
\usepackage{amsfonts}
\usepackage{amssymb}
\usepackage{graphicx}
\usepackage{svg}

\usepackage{subcaption}
\DeclareMathOperator{\relu}{ReLU}

\def\BibTeX{{\rm B\kern-.05em{\sc i\kern-.025em b}\kern-.08em
    T\kern-.1667em\lower.7ex\hbox{E}\kern-.125emX}}

\begin{document}
\title{Using context to adapt to sensor drift}

\author{\IEEEauthorblockN{Jamieson Warner}
\IEEEauthorblockA{\textit{Department of Neuroscience} \\
\textit{The University of Texas at Austin}\\
Austin, TX \\
jamiesonwarner@utexas.edu}
\and
\IEEEauthorblockN{Ashwin Devaraj}
\IEEEauthorblockA{\textit{Department of Computer Science} \\
\textit{The University of Texas at Austin}\\
Austin, TX \\
ashwin.devaraj@utexas.edu}
\and
\IEEEauthorblockN{Risto Miikkulainen}
\IEEEauthorblockA{\textit{Department of Computer Science} \\
\textit{The University of Texas at Austin}\\
Austin, TX \\
risto@cs.utexas.edu}
}

\maketitle

\begin{abstract}
% \textbf{\abstractname.}
Lifelong development allows animals and machines to adapt to changes in the environment as well as in their own systems, such as wear and tear in sensors and actuators. An important use case of such adaptation is industrial odor-sensing. Metal-oxide-based sensors can be used to detect gaseous compounds in the air; however, the gases interact with the sensors, causing their responses to change over time in a process called sensor drift. Sensor drift is irreversible and requires frequent recalibration with additional data. This paper demonstrates that an adaptive system that represents the drift as context for the skill of odor sensing achieves the same goal automatically. After it is trained on the history of changes, a neural network predicts future contexts, allowing the context+skill sensing system to adapt to sensor drift. Evaluated on an industrial dataset of gas-sensor drift, the approach performed better than standard drift-naive and ensembling methods. In this way, the context+skill system emulates the natural ability of animal olfaction systems to adapt to a changing world, and demonstrates how it can be effective in real-world applications.
% By reducing the effect that sensor drift has on classification accuracy, context-based models may be used to extend the effective lifetime of gas identification systems in practical settings.
\end{abstract}


\begin{figure*}[htbp]
	% \centering
	\frame{\includegraphics{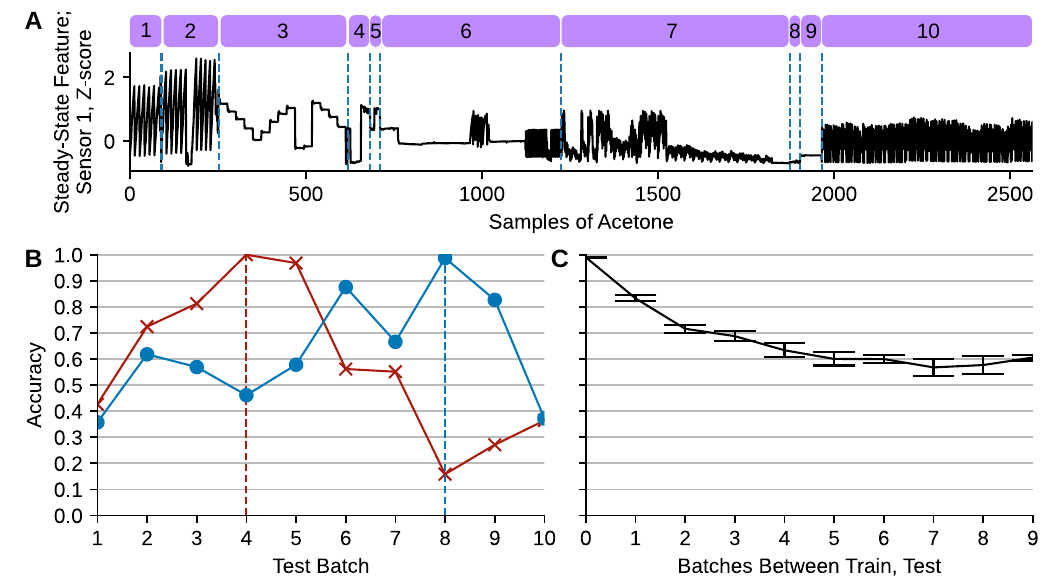}}
	\caption{\textbf{Recognizing odor despite sensor drift.} (\textbf{A.}) The first sensor steady-state response is plotted for all samples of the gas acetone. The data collection periods, called batches, are labeled 1-10 and separated by vertical dashed lines. There is variability within as well as between batches. (\textbf{B.}) For thirty trials, two feedforward networks are trained using a single batch of training data indicated by the vertical dashed lines and colors. The mean classification accuracies evaluated on each batch are shown. The variability of response to a single odor poses a challenge to odor recognition. (\textbf{C.}) Performance degrades as a function of the time between the training batch and the evaluation batch. Mean accuracy along with 95\% confidence interval is shown as a function of the absolute distance between the training batch and the evaluation batch. The further away the testing batch is from the training batch, the lower the generalization accuracy becomes.}
\end{figure*}
\section{Introduction}

Natural systems need to adapt to a changing world continuously; seasons change, food sources and shelter opportunities vary, cooperation and competition with other animals evolves over time. Moreover, their embodiment also changes over their lifetime. Young animals experience a period of growth where their size increases multiple times; in old age, they become less agile and their sensors less accurate. Yet natural systems are remarkably robust against such variation, allowing them to survive and thrive despite the changes.

It is common to try to avoid such changes in artificial agents, machines, and industrial processes. When something changes, the entire system is taken offline and modified to fit the new situation. This process is costly and disruptive; adaptation similar to that in nature might make such systems more reliable and long-term, and thus cheaper to operate.

Sensor drift in industrial processes is one such use case. For example, sensing gases in the environment is mostly tasked to metal oxide-based sensors, chosen for their low cost and ease of use \cite{dickinson_current_1998,barsan_metal_2007}. An array of sensors with variable selectivities, coupled with a pattern recognition algorithm, readily recognizes a broad range of odors. The arrangement is called an artificial nose since it resembles the multiplicity of sensory neuron types in the nasal epithelium. However, while metal oxide-based sensors are economical and flexible, they are unstable over time. Changes to the response properties of sensors make it difficult to detect and identify odors in the long term, and sensors have to be recalibrated to compensate \cite{marco_signal_2012}.  Recalibration requires collecting and labeling new samples, which is costly because a skilled operator is needed, and challenging because the experimental conditions need to be controlled precisely \cite{marco_signal_2012}. Recalibrating a model with unlabeled examples, called semisupervised learning, is a possible alternative but difficult to establish in practice.

An alternative approach is to emulate adaptation in natural sensor systems. The system expects and automatically adapts to sensor drift, and is thus able to maintain its accuracy for a long time. In this manner, the lifetime of sensor systems can be extended without recalibration.

More specifically, natural odors consist of complex and variable mixtures of molecules present at variable concentrations \cite{thomas-danguin_perception_2014}. Sensor variance arises from environmental dynamics of temperature, humidity, and background chemicals, all contributing to concept drift \cite{vito_pattern_2016}, as well as sensor drift arising from modification of the sensing device. The hard problem of olfaction in nature calls for the learning of new odor assocations \cite{imam_rapid_2019}. In an attempt to capture much of this complexity, Vergara et al. \cite{vergara_chemical_2012} developed a publicly available benchmark dataset demonstrating sensor drift over a period of 36 months. This dataset offers a controlled testbed for sensor drift mitigation algorithms and thus defines the scope of this paper.

This paper builds upon previous work with this dataset \cite{vergara_chemical_2012}, which used support vector machine (SVM) ensembles. First, their approach is extended to a modern version of feedforward artificial neural networks (NNs) \cite{pardo_remarks_2004}. Context-based learning is then introduced to utilize sequential structure across batches of data. The context model has two parts: (1) a recurrent context layer, which encodes classification-relevant properties of previously seen data, and (2) a feedforward layer, which integrates the context with the current odor stimulus to generate an odor-class prediction. The results indicate improvement from two sources: The use of neural networks in place of SVMs, and the use of context, particularly in cases where a substantial number of context sequences are available for training. Thus, emulation of adaptation in natural systems leads to an approach that can make a difference in real-world applications.

\section{Related Work}

The context+skill approach used in this paper builds on the substantial body of work on training models in nonstationary environments (Section \ref{section:review-sensor-drift}). It also draws inspiration from models of context-aware odor processing in biological systems (Section \ref{section:review-biological-olfaction}).

\subsubsection{Nonstationary environments}
\label{section:review-sensor-drift}
Machine learning applications frequently deal with data-generating processes that change over time. Applications in such nonstationary environments include power use forecasting, recommendation systems, and environmental sensors \cite{ditzler_learning_2015}. Semisupervised learning, which has received a lot of attention in the sensor community, is characterised by the combined use of easily attainable unlabeled data in addition to the initial labeled dataset \cite{de_vito_semi-supervised_2012,liu_drift_2014,zhang_novel_2016}. Extreme learning machines are also frequently deployed in these settings to efficiently reconfigure neural networks based on the new data \cite{zhang_domain_2015, das_gas_2020, wang_sensor_2021}. Within the standard backpropagation framework, ensembles have been used successfully in this setting; therefore that is what we compare with in this paper \cite{vergara_chemical_2012}.

\subsubsection{Use of odor context in the mammalian olfactory system}
\label{section:review-biological-olfaction}
Biology frequently deals with drift \cite{rule_causes_2019}. For instance olfactory systems are constantly adapting, predominantly through feedback mechanisms. This section details some such models from computer science and neuroscience \cite{marco_recent_2009}. One example is the KIII model, a dynamic network resembling the olfactory bulb and feedforward and feedback connections to and from the higher-level anterior olfactory nucleus and piriform cortex \cite{fu_pattern_2007}. Applied to an odor recognition task, KIII performed better than an artificial neural network under sensor drift and variable concentrations, a similar setting to the one in this paper.

One prominent feature of the mammalian olfactory system is feedback connections to the olfactory bulb from higher-level processing regions. Activity in the olfactory bulb is heavily influenced by behavioral and value-based information \cite{kay_odor-_1999}, and in fact, the bulb receives more neural projections from higher-level regions than from the nose \cite{shipley_functional_1996}. In computational modeling, this principle has been taken into account by the piriform cortical region that recognizes familiar background odors through associative memory \cite{adams_top-down_2019}. It projects this information to the olfactory bulb to improve odor recognition when there are background odors. Following this same principle, the neural network classifier in this paper integrates context that is outside the immediate input signal.

\section{Methods}
This section describes the task of generalization of odor classification under sensor drift and defines several classifier models: the SVM ensemble, neural network ensemble, skill neural network, and context+skill neural network.

\subsection{Dataset description}
Experiments in this paper used the gas sensor drift array dataset \cite{vergara_chemical_2012}. The data consists of 10 sequential collection periods, called \textbf{batches}. Every batch contains between $161$ to $3{,}600$ samples, and each sample is represented by a 128-dimensional feature vector; 8 features each from 16 metal oxide-based gas sensors. These features summarizing the time series sensor responses are the raw and normalized steady-state features and the exponential moving average of the increasing and decaying transients taken at three different alpha values. The experiments used six gases, ammonia, acetaldehyde, acetone, ethylene, ethanol, and toluene, presented in arbitrary order and at variable concentrations. Chemical interferents were also presented to the sensors between batches, and the time between presentations varied, both of which contributed to further sensor variability. The dataset thus exemplifies sensor variance due to contamination and variable odor concentration in a controlled setting.

Two processing steps were applied to the data used by all models included in this paper. The first preprocessing step was to remove all samples taken for gas 6, toluene, because there were no toluene samples in batches 3, 4, and 5. Data was too incomplete for drawing meaningful conclusions. Also, with such data missing it was not possible to construct contexts from odor samples from each class in previous batches. The second preprocessing step normalized each feature so that all values corresponding to any feature dimension of the 128 total have zero mean and unit variance as is standard practice in deep learning.

\subsection{Support vector machines}

The first model in this domain \cite{vergara_chemical_2012} employed SVMs with one-vs-one comparisons between all classes. SVM classifiers project the data into a higher dimensional space using a kernel function and then find a linear separator in that space that gives the largest distance between the two classes compared while minimizing the number of incorrectly labeled samples. In the one-vs-one design, several SVMs are trained to discriminate between each pair of classes, and the final multiclass prediction is given by the class with the majority of votes.

In order to improve performance, Vergara et al. \cite{vergara_chemical_2012} employed an ensemble technique on the SVM classifiers (Fig. 2B). The same technique was reimplemented and tested on the modified dataset in this paper. The ensemble meant to generalize to batch $T$ was constructed by training a collection of single-batch classifiers, so that for every batch 1 through $T-1$, a model is trained using that batch as the training set. Then, each model is assigned a weight $\beta_i$ equal to its classification accuracy on batch $T-1$, under the assumption that the most similar batch to batch $T$ will be batch $T-1$. To classify a sample from batch $T$ using the weighted collection of single-batch classifiers, a weighted voting procedure is used. The output of the ensemble is equal to the weighted sum of all classifiers' outputs.

The experiments were based on the Scikit-learn Python library \cite{pedregosa_scikit-learn_2011}, which implements its SVM classifiers using LibSVM \cite{chang_libsvm_2011}. The SVMs use a radial basis function (RBF) kernel. The two hyperparameters $C$, which is the penalty cost of incorrectly classifying a training sample, and $\gamma$, which determines the steepness of the RBF function, were determined by 10-fold cross-validation. In this scheme, the training batch is partitioned into 10  sets, called folds. The accuracy of each hyperparameter configuration in the range $C \in \{ 2^{-5}, 2^{-4}, 2^{-3}, \ldots, 2^{10} \}$ and $\gamma \in \{ 2^{-10}, 2^{-9}, 2^{-8}, \ldots, 2^{5} \}$ was evaluated by the average accuracy over ten folds, taken by training a model on nine folds and calculating its accuracy on the remaining fold. The selected hyperparameters maximize the mean accuracy over the evaluated folds.

\begin{figure*}
	\centering
	\frame{\includegraphics{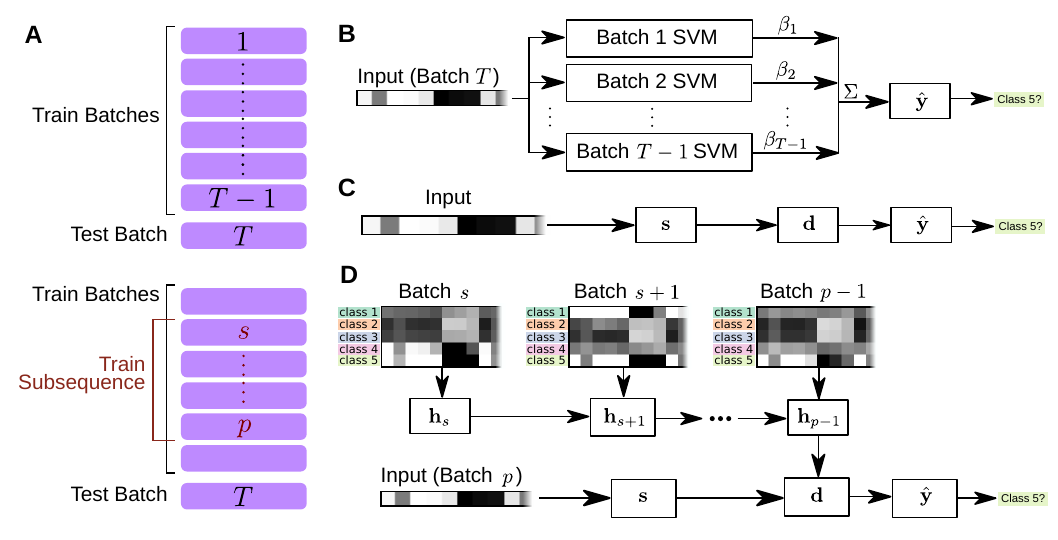}}
	\caption{\textbf{Neural network architectures.} (\textbf{A}.) The batches used for training and testing illustrate the training procedure. The first $T-1$ batches are used for training, while the next unseen batch $T$ is used for evaluation. When training the context network, subsequences of the training data are selected to be processed recurrently, indicated by the labels $s$ through $p$. In all cases, training data is obtained only from the first $T-1$ batches of data. (\textbf{B}.) A feature vector is input to a collection of SVMs, one trained on each prior batch. Each SVM output is weighted by its corresponding coefficient, $\beta$, and the weighted sum of the output class predictions is taken to be the output, $\hat{\mathbf{y}}$, of the ensemble. (\textbf{C}.) A schematic of the skill model shows feedforward progression of input through two hidden layers $\mathbf{s}$ and $\mathbf{d}$ followed by the output layer $\hat{\mathbf{y}}$. (\textbf{D}.) A schematic of the context+skill model introduces a sequential processing of prior samples as a separate processing pathway. For each context batch from $s$ through $p-1$, one sample per odor class is chosen as a representative. The context information is then utilized by the ``decision-making'' layer $\mathbf{d}$ and is thus integrated into the feedforward pathway.}
\end{figure*}

\subsection{Artificial neural networks}

While SVMs are standard machine learning, NNs have recently turned out more powerful, so the first step is to use them on this task instead of SVMs. In the classification task, the networks are evaluated by the similarity between the odor class label (1-5) and the network's output class label prediction given the unlabeled odor features. The output layer of the neural network is a five-dimensional vector of class scores, which represent the network's confidence attributed to each odor class label. The cross-entropy loss function combines the predicted class scores and the true odor label to form the network training signal. Assume that $\hat{\mathbf{y}}$ is the five-dimensional unnormalized class score vector, that $\hat{y}_i$ represents the score given to class $i$, and that $c$ is the identity of the true class label. Then the loss function is given
\begin{equation}
\mathcal{L} = - \hat{y}_c + \log \left( \sum_{i=1}^5 \exp(\hat{y}_i) \right).
\end{equation}

All neural networks in this section were trained using stochastic gradient descent with momentum \cite{rumelhart_learning_1986} on the loss function $\mathcal{L}$. The learning rate was set to $10^{-3}$ and the momentum factor to $0.9$. Networks were trained for 200 epochs under a weight decay factor of $10^{-2}$, which guards against overfitting the model to the training data \cite{pardo_remarks_2004}. The initial weights of each layer are sampled from a Gaussian distribution with zero mean and variance equal to the reciprocal of the number of units in the layer. The neural networks were implemented using the PyTorch backpropagation library \cite{paszke_pytorch_2019}.

Two neural network architectures are evaluated: a skill-only network as the baseline and the context+skill network that models the drift separately.

\subsubsection{The skill model}
The skill network approach incorporates all available data into a single training set, disregarding the sequential structure between batches of the dataset. For each batch $T$, a network was trained using batches $1$ through $T-1$ as the training set and evaluated on batch $T$.

The network is diagrammed in Figure 2C. It is input the 128-dimensional feature vector, $\mathbf{x}$, and calculates class scores, $\hat{\mathbf{y}}$, using two hidden layers: the 50-unit ``skill" layer, $\mathbf{s}$, and the 20-unit ``decision-making" layer, $\mathbf{d}$. Assume that $\mathbf{W}_{xs}$, $\mathbf{W}_{sd}$, and $\mathbf{W}_{dy}$ are matrices of dimensions $50\times 128$, $20\times 50$, and $20\times 5$, respectively, and that $\mathbf{b}_s, \mathbf{b}_d,$ and $\mathbf{b}_y$ are bias vectors of dimensions $50, 20,$ and $5$, respectively. Define $\relu$ to be the rectified linear activation function ($\relu(\mathbf{x})_i = \max(0, x_i)$). Then, the feedforward NN model is given by
\begin{equation}
\begin{split}
\mathbf{s}=\relu(\mathbf{W}_{xs}\cdot\mathbf{x}+\mathbf{b}_s), \\
\mathbf{d}=\relu(\mathbf{W}_{sd}\cdot\mathbf{s}+\mathbf{b}_d), \\
\hat{\mathbf{y}}=\mathbf{W}_{dy}\cdot\mathbf{d}+\mathbf{b}_y.
\end{split}
\end{equation}

This paper also presents the NN ensemble created in the same way as with SVMs. In the NN ensemble, $T-1$ skill networks are trained using one batch each for training. Each model is assigned a weight $\beta_i$ equal to its accuracy on batch $T-1$. The weighted sum of the model class scores is the ensemble class prediction. The model is then tasked to classify samples from batch $T$.

\subsubsection{The context+skill model}
The context+skill NN model builds on the skill NN model by adding a recurrent processing pathway (Fig. 2D). Before classifying an unlabeled sample, the recurrent pathway processes a sequence of labeled samples from the preceding batches to generate a context representation, which is fed into the skill processing layer. The recurrent layers are modified via backpropagation through time, and, in this manner, the recurrent pathway learns to generate representations that support classification. The context system thus transforms samples of recently seen odors into a representation that helps classification on the next time period. This approach is similar to the context+skill technique for opponent modeling and enhanced extrapolation in games \cite{li:gecco18,tutum:cog21}; the main difference is that in prior work the approach was based on neuroevolution of agent behavior, whereas in this paper it is implemented via backpropagation to generalize classification performance.

The context pathway is based on a recurrent neural network (RNN) approach. It reuses weights and biases across the steps of a sequence and can thus process variable-length sequences. The alternative was to use a long-short term memory (LSTM), which employs gating variables to better remember information in long sequences \cite{hochreiter_long_1997}. However, in preliminary experiments LSTM did not improve generalization accuracy significantly ($p\geq 0.05$, one-sided t-test blocked by batch), presumably because the sequences were relatively short (nine steps or less). The simple RNN was therefore used in the experiments presented in the paper.

During training, variable-length context sequences are sampled from the available training data (Fig. 2A). For each unlabeled sample in the training set, a context sequence is sampled from the preceding batches. If the sample is taken from batch $p$, then the sequence of context begins at batch $s$, sampled uniformly from $s \in \{ 1, 2, \ldots, p-1 \}$, and the sequence ends at batch $p-1$. During training, all data is taken from the training set; i.e., $p < T$. At test time when evaluating on batch $T$, all batches $1$ through $T-1$ are included for context; i.e., $s=1$, and $p=T$. At each step in the context sequence, labeled odors are sampled for each odor class to be provided as input to the context network. Denote $\mathbf{x}_\mathrm{c}^{t,z}$ to be the feature vector sampled from batch $t$ and odor class $z$. The concatenation over all five classes in order forms a single vector for batch $t$ denoted $\mathbf{x}_\mathrm{c}^{t} = (\mathbf{x}_\mathrm{c}^{t,1}, \ldots, \mathbf{x}_\mathrm{c}^{t,5}) $, which is the input to the RNN. The RNN iterates from batch $s$ to batch $p-1$ to form the hidden state by the recursive formula
\begin{equation}
\begin{split}
\mathbf{h}_s = \tanh(\mathbf{W}_{xh} \cdot \mathbf{x}_\mathrm{c}^{s} + \mathbf{b}_h),\\
\mathbf{h}_t = \tanh(\mathbf{W}_{hh}\cdot \mathbf{h}_{t-1} + \mathbf{W}_{xh} \cdot \mathbf{x}_\mathrm{c}^{t} +  \mathbf{b}_h).
\end{split}
\end{equation}

The output, $\mathbf{h}_{p-1}$,  of the recurrent pathway is integrated with the feedforward pathway. First, the unlabeled target sample $\mathbf{x}$ is processed in a 50-unit ``skill" layer, $\mathbf{s}$. Then, the 50-dimensional representation of $\mathbf{x}$ and the 10-dimensional context vector are combined in a 20-unit ``decision-making" layer, $\mathbf{d}$.  The class scores $\hat{\mathbf{y}}$ are a linear readout of $\mathbf{d}$. The entire model is thus given as
\begin{equation}
\begin{split}
\mathbf{s}=\relu(\mathbf{W}_{xs}\cdot\mathbf{x}+\mathbf{b}_{s}), \\
\mathbf{d}=\relu(\mathbf{W}_{sd}\cdot\mathbf{s} + \mathbf{W}_{hd}\cdot\mathbf{h}_{p-1}+\mathbf{b}_{d} ), \\
\hat{\mathbf{y}}=\mathbf{W}_{dy}\cdot\mathbf{d}+\mathbf{b}_{y}.
\end{split}
\end{equation}

The context processing pathway utilizes the sequential structure of the dataset via recurrent processing. This pathway is incorporated with a feedforward component to define the context+skill model as described above.

\begin{table*}[t]

% \captionsetup{justification=centering,font=sf}
% \captionsetup{justification=centering, labelsep=newline}
\centering
% \caption{Mean Generalization Accuracy}
\setlength{\tabcolsep}{1em}\begin{tabular}{lrrrrrrrrr}
& \multicolumn{6}{l}{Batch }\\
\cline{2-10}
 Model        &       3 &       4 &       5 &       6 &       7 &       8 &       9 &      10 &   $\mu$\\
\hline
 Feedforward   & 0.881          & 0.875          & 0.974          & \textbf{0.959} & 0.792          & 0.839          & 0.896          & 0.737          & 0.869          \\
 Feedforward+Context      & 0.882          & 0.869          & 0.975          & 0.947          & 0.820          & 0.864          & \textbf{0.939} & \textbf{0.763} & \textbf{0.882} \\
 Feedforward NN Ensemble  & \textbf{0.921} & \textbf{0.904} & 0.979 & 0.903          & 0.777          & 0.679          & 0.864          & 0.693          & 0.840          \\
 SVM Ensemble & 0.698          & 0.777          & 0.631          & 0.900          & 0.823 & \textbf{0.934} & 0.794          & 0.551          & 0.764          \\
\hline
\end{tabular}
\caption{\textbf{Mean generalization accuracy.} Listed is the classification accuracy (correct / total) of various models evaluated on the unseen testing data, i.e., batch $T$. The values represent the average accuracy over 30 trials. The final column lists the mean of the values for batches 3 through 10. A bolded value is significantly greater than the others in the same column ($p < 0.05$, two-sided pairwise t-test with correction for unequal variances).}
\end{table*}

\begin{figure*}
	\frame{\includegraphics{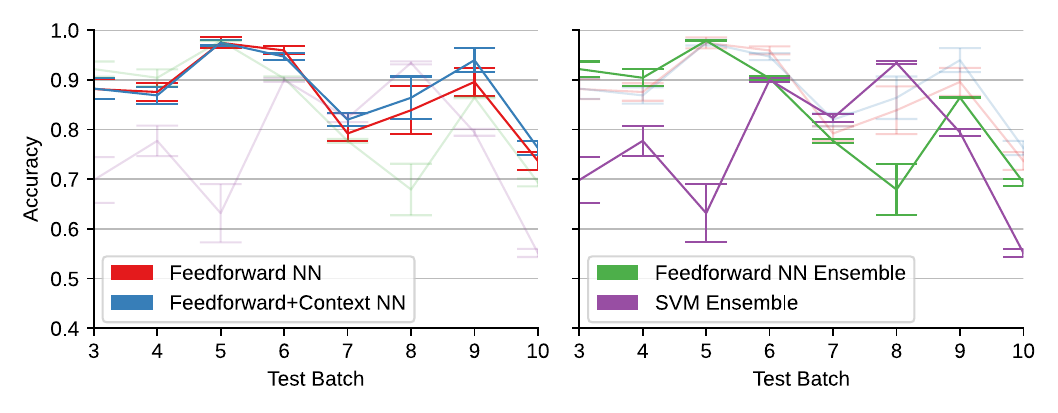}}
	\caption{\textbf{Generalization accuracy.} The generalization accuracy of each model was evaluated on batch $T$. For each model type and every batch, 30 models were trained. The line represents the average over the 30 trials, and the error bar is the 95\% confidence interval. (\textbf{A}.) The skill and context+skill models are shown with the other models faded out. The context most contributes to performance in the later batches, which offer the longest context sequences. (\textbf{B}.) The SVM ensemble and NN ensemble models are shown with the results from (A) are shown faded out. Both ensemble models are variable in performance between batches. }
\end{figure*}

\section{Experiments}
A series of experiments were carried out to evaluate the following hypotheses:
\begin{enumerate}
\item Sensor drift has a sequential structure.
\item Neural network ensembles offer better generalization properties than do SVM ensembles.
\item The inclusion of context improves the generalization performance of neural networks.
\end{enumerate}

\subsection{Drift demonstration}
First, the effect of sensor drift on classification accuracy is demonstrated using classifiers trained on a single batch. For each batch $1$ through $10$, a feedforward model was trained on that batch. Training of a new model was repeated 30 times on each batch. The accuracy of all classifiers were evaluated on every batch. Networks trained on batches 2, 4, 6, and 8 were plotted (Fig. 1B).  The accuracy data was reformulated as a function of the distance between the train batch and the test batch (Fig. 1C). As expected, the accuracy decreases with the time gap between training and testing, demonstrating that, indeed, sensor drift progresses over time.

\subsection{SVM vs.\ NN ensemble models}

The second comparison is between the weighted ensembles of SVMs, i.e., the state of the art \cite{vergara_chemical_2012}, and the weighted ensembles of neural networks. For each batch, an SVM and a neural network were trained with that batch as the training set. Weighted ensembles were constructed for each batch $T$ by assigning weights to the models trained on batches 1 through $T-1$. Training of a new ensemble and evaluation were repeated for thirty trials. The generalization accuracy classifying samples in batch $T$ was reported for the neural network ensemble (Fig. 3B, Table 1, \textit{NN Ensemble}) and for the SVM ensemble (Fig. 3B, Table 1, \textit{SVM Ensemble}). The neural network ensemble model had a significantly greater mean generalization accuracy than the SVM ensemble model ($p < 0.05$, two-sided t-test blocked by batch), and the neural network ensemble model achieved the highest generalization accuracy among all models on batches 3 and 4 ($p < 0.05$, pairwise two-sided t-test). The results indicate that while NN ensembles outperform SVM ensembles on average, there is significant batch to batch variability.

\subsection{Including context}
For each batch $T$ from 3 through 10, the batches $1, 2, \ldots, T-1$ were used to train skill NN and context+skill NN models for 30 random initializations of the starting weights. The accuracy was measured classifying examples from batch $T$ (Fig. 3A, Table 1, \textit{Skill NN} and \textit{Context+Skill NN}). The context models achieved a greater average accuracy, computed as the mean over all batches tested of the average accuracy in that batch ($p<0.05$, two-sided t-test blocked by batch). In batch 6, the skill NN outperformed the context+skill NN, while the context+skill NN achieved greater performance in batches 7, 9, and 10 (two-sided t-tests, $p<0.05$).

\section{Discussion}
The purpose of this study was to demonstrate that explicit representation of context can allow a classification system to adapt to sensor drift. Several gas classifier models were placed in a setting with progressive sensor drift and were evaluated on samples from future contexts. This task reflects the practical goal to deploy an artificial nose in a dynamic environment without recalibration.

First, extending the previous SVM ensemble model \cite{vergara_chemical_2012}, a similar ensemble was formed from neural networks. The accuracy improvements suggested that neural networks are indeed more accurate than SVMs in this task.

Second, skill NN and context+skill NN models were compared. The context-based network extracts features from preceding batches in sequence in order to model how the sensors drift over time. When added to the feedforward NN representation, such contextual information resulted in improved ability to compensate for sensor drift. This benefit was larger in later batches where the drift was the largest and where there was a longer context to use as a basis for the adaptation.

The approach is biologically motivated in that in many sensory systems, top-down feedback outnumbers bottom-up connections \cite{fu_pattern_2007}. Such a structure is well suited to take advantage of context representations, allowing the bottom-up skill performance to be modulated by the top-down context.

While context did introduce more parameters to the model ($7{,}575$ parameters without context versus $14{,}315$ including context), the model is still very small compared to most neural network models, and is trainable in a few hours on a CPU. When units were added to the ``skill" layer of the feedforward NN model until the total number of parameters reached $14{,}429$, the larger model was not significantly better ($p\geq 0.05$, one-sided t-test blocked by batch). This reinforces the idea that the benefit may be attributed to context, and not to the size of the network.

The estimation of context by learned temporal patterns should be most effective when the environment results in recurring or cyclical patterns, such as in cyclical variations of temperature and humidity and regular patterns of human behavior generating interferents. In such cases, the recurrent pathway can identify useful patterns analagously to how cortical regions help the olfactory bulb filter out previously seen background information \cite{adams_top-down_2019}. A context-based approach will be applied to longer-timescale data and to environments with cyclical patterns.

The current design of the context-based network relies on labeled data because the odor samples for a given class are presented as ordered input to the context layer. However, the model can be modified to be trained on unlabeled data, simply by allowing arbitrary data samples as input to the context layer. This design introduces variation in training inputs, which makes it harder to learn consistent context patterns. For this task, semisupervised learning techniques, such as self-labeled samples, may help. If the context layer can process unlabeled data, then it is no longer necessary to include every class in every batch. The full six-gas sensor drift dataset can be used, as well as other unbalanced and therefore realistic datasets.

\section{Conclusion}
While natural systems cope with changing environments and embodiments well, they form a serious challenge for artificial systems. For instance, to stay reliable over time, gas sensing systems must be continuously recalibrated to stay accurate in a changing physical environment. Drawing motivation from nature, this paper introduced an approach based on continual adaptation. A recurrent neural network uses a sequence of previously seen gas recordings to form a representation of the current state of the sensors. It then modulates the skill of odor recognition with this context, allowing the system to adapt to sensor drift.  Context models can thus play a useful role in lifelong adaptation to changing environments in artificial systems.

\section{Acknowledgments}
This research was made possible by assistance from the Defense Advanced Research Projects Agency (DARPA) Lifelong Learning Machines program under grant HR0011-18-2-0024 and from the National Institutes of Health (NIH) grant R01DC020653.

\bibliographystyle{unsrt}
\bibliography{SensorDrift}

\end{document}